\documentclass[twocolumn]{aastex62}
\usepackage{aas_macros}
\usepackage{graphicx}
\usepackage{epstopdf}
\usepackage{mathrsfs}
\usepackage{hyperref}
\usepackage{amsmath}
\usepackage{xspace}
\usepackage{natbib}
\usepackage{amsmath}
\usepackage{natbib}

\newcommand{\kms}{\ensuremath{\mathrm{km s}^{-1}}\xspace}

\begin{document}
\title{Revealing the Formation of the Milky Way Nuclear Star Cluster via Chemo-Dynamical Modeling}
\author{Tuan Do}
\affiliation{UCLA Galactic Center Group, Physics and Astronomy Department, UCLA, Los Angeles, CA 90095-1547}
\author{Gregory David Martinez}
\affiliation{UCLA Galactic Center Group, Physics and Astronomy Department, UCLA, Los Angeles, CA 90095-1547}
\author{Wolfgang Kerzendorf}
\affiliation{Department of Physics and Astronomy, Michigan State University, East Lansing, MI 48824, USA}
\affiliation{Department of Computational Mathematics, Science, and Engineering, Michigan State University, East Lansing, MI 48824, USA}
\author{Anja Feldmeier-Krause}
\affiliation{The Department of Astronomy and Astrophysics, The University of Chicago, 5640 S. Ellis Ave, Chicago, IL 60637, USA}
\author{Manuel Arca Sedda}
\affiliation{Astronomisches Rechen Institut, Zentrum f\"ur Astronomie der Universit\"at Heidelberg, D-69120 Heidelberg, Germany}
\author{Nadine Neumayer}
\affiliation{Max-Planck-Institute for Astronomy, Heidelberg, Germany}
\author{Alessia Gualandris}
\affiliation{Department of Physics, University of Surrey, Guildford GU2 7XH, UK}
\correspondingauthor{Tuan Do}
\email{tdo@astro.ucla.edu}

\begin{abstract}
The Milky Way nuclear star cluster (MW NSC) has been used as a template to understand the origin and evolution of galactic nuclei and the interaction of nuclear star clusters with supermassive black holes. It is the only nuclear star cluster with a supermassive black hole where we can resolve individual stars to measure their kinematics and metal abundance to reconstruct its formation history. Here, we present results of the first chemo-dynamical model of the inner 1 pc of the MW NSC using metallicity and radial velocity data from the KMOS spectrograph on the Very Large Telescope. We find evidence for two kinematically and chemically distinct components in this region. The majority of the stars belong to a previously-known super-solar metallicity component with a rotation axis perpendicular to the Galactic plane. However, we identify a new kinematically distinct sub-solar metallicity component which contains about 7\% of the stars and appears to be rotating faster than the main component with a rotation axis that may be misaligned. This second component may be evidence for an infalling star cluster or remnants of a dwarf galaxy, merging with the MW NSC. These measurements show that the combination of chemical abundances with kinematics is a promising method to directly study the MW NSC's origin and evolution. 
\end{abstract}

\keywords{Galaxy: center --- stars: late-type --- techniques: spectroscopic}

\section{Introduction}

The center of the Milky Way (MW) offers us a laboratory for the study of the formation and evolution of galactic nuclei. At 8 kpc from Earth \citep{2019A&A...625L..10G,2019Sci...365..664D}, the Galactic center hosts the closest example of a nuclear star cluster (NSC) with a supermassive black hole. Its proximity offers us the opportunity to measure the physical properties of individual stars, such as their motion and chemical composition, to reconstruct the NSC's formation history. This will allow us to test models for the formation of galactic nuclei and their chemical enrichment history \citep[e.g.][]{2019ApJ...886...57A,2020arXiv200207814A}. The MW NSC is the most massive and densest star cluster in the Galaxy with multiple stellar populations \citep[e.g.][]{2011ApJ...741..108P,2013ApJ...764..154D,2013ApJ...764..155L} and a wide range of metallicities \citep[e.g.][]{2015ApJ...809..143D,2015A&A...573A..14R,2017MNRAS.464..194F,2017AJ....154..239R,2018ApJ...855L...5D,2020arXiv200305998F}. These properties suggest a complex formation history, which may be revealed by combining the different sources of data into a coherent model of the cluster. 

Here, we present the first chemo-dynamical model of the MW NSC. By combining both the metallicity measurements and the kinematics, our goal is to search for signatures of its formation history. For example, if the cluster is built by the infall of star clusters or dwarf galaxies \citep[e.g.][]{2008ApJ...681.1136C,2012ApJ...750..111A,2015ApJ...806..220A,2018MNRAS.479..900A,2020arXiv200103626N}, there may be distinct chemical and kinematic components depending on the age of the infall. If the entire cluster is formed \textit{in-situ}, we may detect more spatially uniform kinematic and chemical signatures. We introduce the observations in Section \ref{sec:obs}. We present the methodology and the results of the modeling in Section \ref{sec:results}. Finally, we discuss the evidence for a chemical-kinematic sub-structure and its implications in Section \ref{sec:discussion}.

\section{Observations and Data}
\label{sec:obs}
The data for this work were presented and described in \citet{2017MNRAS.464..194F}. In summary, these data are observations from the KMOS integral-field spectrograph with a field of view of $\sim60\arcsec\times40\arcsec$ ($2.4\times1.6$ pc in projection) centered on Sgr A*, the radio-source associated with the supermassive black hole at the Galactic center. The spectra are taken in the K-band (1.934 - 2.460 $\mu$m) and a spectral resolution of about R$\sim$ 4000. Stellar parameters such as the effective temperature, metallicity, and radial velocities were measured for each star using StarKit \citep{2015zndo.soft28016K,2018ApJ...855L...5D}. We use the sample of late-type red giants and their measured projected position, metallicity, and radial velocity in this study. 

\section{Methodology and Results}
\label{sec:results}

\subsection{Kinematics and Metallicity}

\begin{figure}[tbh]
 \center
 \includegraphics[width=3.25in]{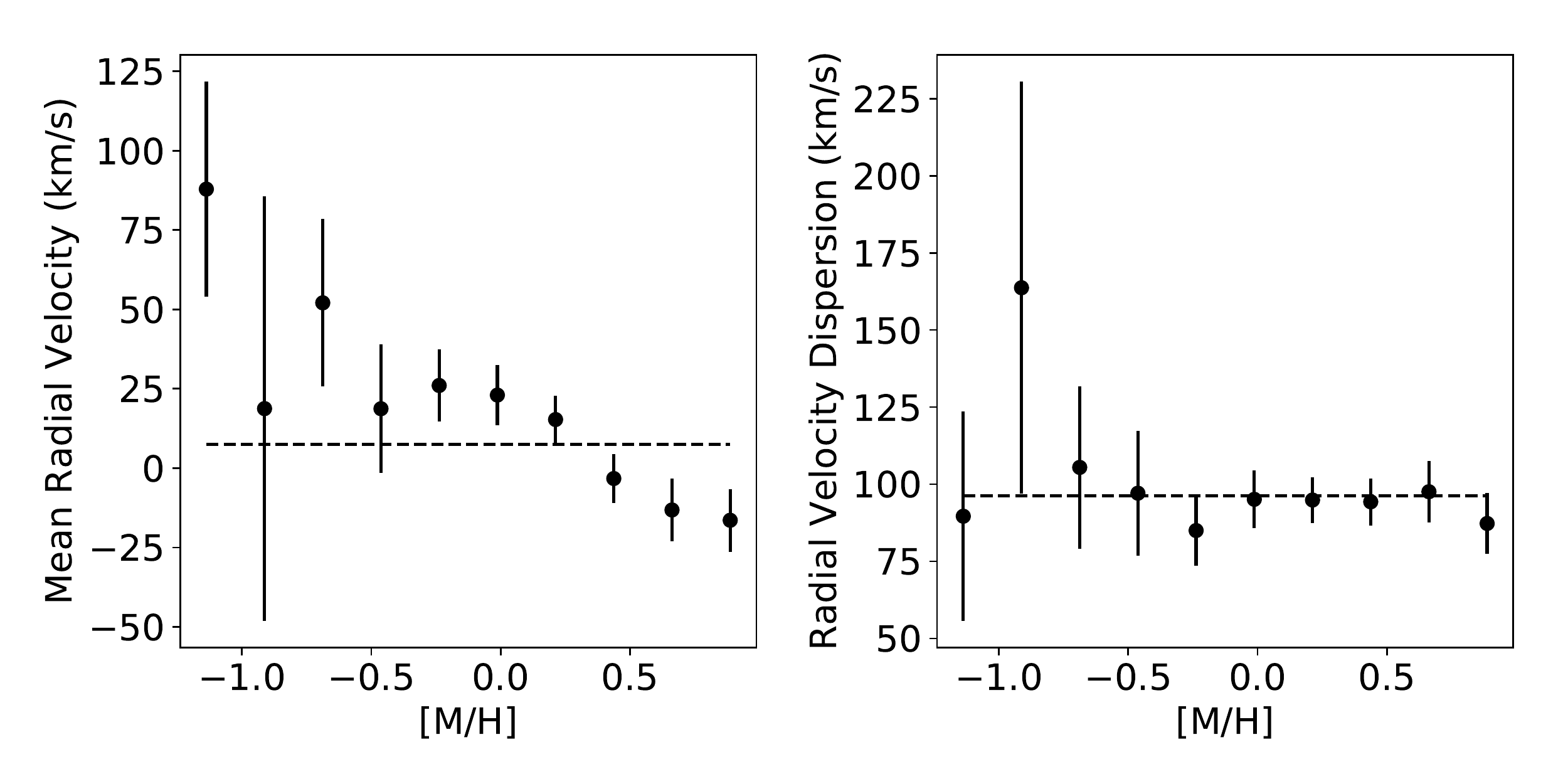}
 \caption{\textbf{Left:} The mean radial velocity of stars as a function of metallicity in our sample. The uncertainties in the mean velocity is the velocity dispersion divided by the square root of the number of stars. \textbf{Right:} The velocity dispersion of stars as a function of metallicity. While the mean radial velocity appears to be correlated with metallicity, the velocity dispersion is not. This suggests that there is a chemical dynamical separation in the nuclear star cluster and that the different metallicity populations are consistent with being located at the Galactic center.}
 \label{fig:rv_mean}
 \end{figure}

Before we model the NSC, we look for a dependence of the dynamics of the cluster on the metallicity of stars. First, we measure the average radial velocity as a function of metallicity. Figure \ref{fig:rv_mean} shows relationships between the mean radial velocity and velocity dispersion with the bulk metallicity of stars. We use bins of 0.25 dex in $[M/H]$ to allow enough stars (between 6 to 150 stars) in each bin to measure the average radial velocity (RV) and velocity dispersion. The uncertainties in each bin are estimated by the standard deviation divided by the square-root of the number of stars. While some of the low-metallicity bins have larger uncertainties, there appears to be a correlation between metallicity and mean velocity. There is no such correlation with velocity dispersion. This suggests that there is a chemo-dynamical separation in the nuclear star cluster and that the different metallicity populations are consistent with being located at the Galactic center (stars not within the sphere of influence of the black hole should show lower velocity dispersion).

\begin{figure*}[tbh]
 \center
 \includegraphics[width=6.75in]{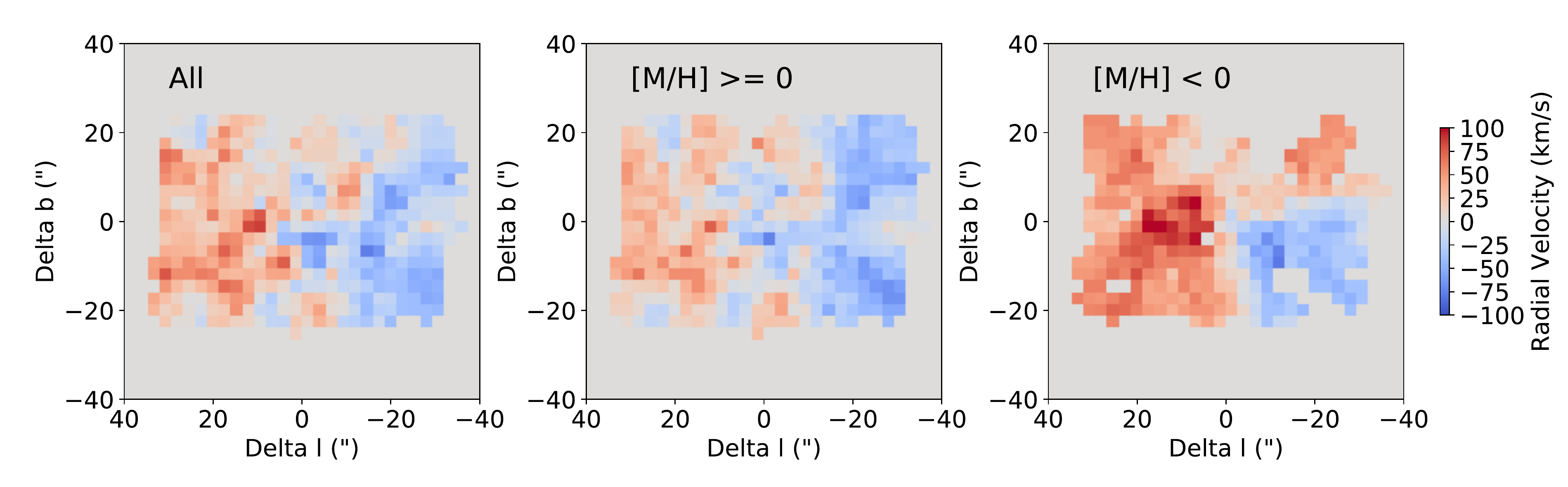}
 \caption{Radial velocity maps of the central $\sim1.5$ pc radius around the supermassive black hole at the Galactic center. These maps are constructed from the average of the radial velocity of nearest 15 neighbor stars at each location. \textbf{Left:} The full sample of KMOS stars. \textbf{Center:} stars with higher than solar metallicity. \textbf{Right:} stars with lower than solar metallicity. The stars above and below solar metallicity appear to show distinctively different kinematic features.  }
 \label{fig:rv_map}
 \end{figure*}

We also find evidence for spatial variations in the mean radial velocity of sub-solar metallicity stars compared to super-solar metallicity stars. In Figure \ref{fig:rv_map}, we show the spatially binned radial velocity map of the $[M/H] < 0$ and the $[M/H] > 0$ populations. The super-solar metallicity population shows rotation with a rotation axis perpendicular to the Galactic plane. The smaller population of sub-solar metallicity stars appears to show stronger rotation, perhaps also mis-aligned compared to the bulk of the stars. In order to better quantify these features, we introduce a simple chemo-dynamical model below. 
 
\subsection{Chemo-dynamical model}

The rich dataset on the MW NSC allows us to model the dynamics of stars of different metallicities simultaneously. We use a chemo-dynamical model to separate possible differences in the dynamics of the stars as a function of metallicity. We use a two-population Gaussian mixture model for the data. In such a model, we can use all the measurements and their uncertainties for each star without the disadvantages of binning the data. For each star, $i$, the probability of observing the star is:
\begin{equation}
P_i(X_i|\theta_1,\theta_2) = fP_1(X_i|\theta_1) + (1-f)P_2(X_i|\theta_2)
\end{equation}
where $P_1$ and $P_2$ are the likelihood of Population 1 and 2 respectively, $X_i$ is the input data for the star, and $\theta_1$ and $\theta_2$ are the model parameters of Population 1 and 2. $f$ is the fraction of stars belonging to Population 1. We model each population with the following properties: (1) metallicity distribution, (2) radial velocity distribution, and (3) rotational velocity. We model the metallicity distribution as a Gaussian:
\begin{equation}
\begin{split}
P([M/H],\sigma_{[M/H]}|m,\sigma_m) \propto \\
\exp\left(\frac{-([M/H] - m)^2}{2(\sigma_{[M/H]}^2 + \sigma_m^2)}\right),
\end{split}
\end{equation}
where $[M/H]$ and $\sigma_{[M/H]}$ are the observed metallicity and its uncertainty and $m$ and $\sigma_m$ are the mean metallicity and intrinsic metallicity dispersion. We model the radial velocity of each population as a solid-body rotator with mean velocity and intrinsic velocity dispersion as projected on the plane of the sky: 
\begin{equation}
\begin{split}
RV &= v_{z,o} + \frac{v_x x}{R} + \frac{v_y y}{R} \\
\sigma^2 &= \sigma_{RV}^2+ \sigma_{v_z}^2 \\
P(x,y,RV,\sigma_{RV}|v_{z},\sigma_{v_z},v_x,v_y) &\propto \exp\left(-(RV - v)^2/2\sigma^2\right),
\end{split}
\end{equation}
where $RV$, $\sigma_{RV}$ are the observed radial velocity and radial velocity uncertainty at a projected distance $x$, $y$ from Sgr A* and a scale radius of $R = 40\arcsec$ (approximately the edge of the field of view). The mean model velocity parameters in each direction are $v_x$, $v_y$ and $v_z$ and $\sigma_{v_z}$ is the intrinsic dispersion along the line of sight. 

In total, we use 13 parameters in our chemical-dynamical model. Each component has 6 parameters $(m,\sigma_m,v_z,\sigma_{v_z},v_x,v_y)$ and one parameter describes the relative fraction ($f$) of the two populations. 

To fit the Gaussian mixture model, we use Bayesian inference. The posterior distribution of the model parameters can be described as:
\begin{equation}
P(\theta|d) = \frac{P(d|\theta)P(\theta)}{P(d)},
\end{equation}
where $\theta$ are the model parameters (described above) and $d$ is the data (projected position of the stars, $x,y$, and radial velocities $RV, \sigma_{RV}$). 
To sample the posterior, we use nested-sampling with the MultiNest algorithm \citep{2009MNRAS.398.1601F}. We give the central confidence interval for these parameters in Table 1.

\begin{deluxetable}{lcc}[]
\tablecolumns{3}
\centering
\tablecaption{Best-fit model parameters}
\tablehead{\colhead{Parameter} & \colhead{Population 1 } & \colhead{Population 2} }
\startdata
Mean Velocity ($v_z$, \kms) & $    0.41^{+4.15}_{-4.55}$ & $   43.48^{+23.87}_{-22.21}$ \\
Velocity Dispersion ($\sigma_{v_z}$, \kms) & $   91.84^{+2.71}_{-2.71}$ & $  108.96^{+19.37}_{-14.37}$ \\
Vx ($v_x$, \kms) & $   48.10^{+8.75}_{-8.87}$ & $   77.45^{+58.69}_{-48.73}$ \\
Vy ($v_y$, \kms) & $    0.06^{+13.76}_{-13.64}$ & $   19.39^{+72.99}_{-75.65}$ \\
Mean Metallicity ($m$, dex) & $    0.33^{+0.03}_{-0.02}$ & $   -0.54^{+0.29}_{-0.22}$ \\
Metallicity Spread ($\sigma_m$, dex) & $    0.23^{+0.02}_{-0.03}$ & $    0.29^{+0.12}_{-0.16}$ \\
Fraction Pop2/Pop1 ($f$) &  & $    0.07^{+0.07}_{-0.03}$ \\
\enddata
\label{tab:result}
\end{deluxetable}

\begin{figure}[tbh]
 \center
 \includegraphics[width=3.25in]{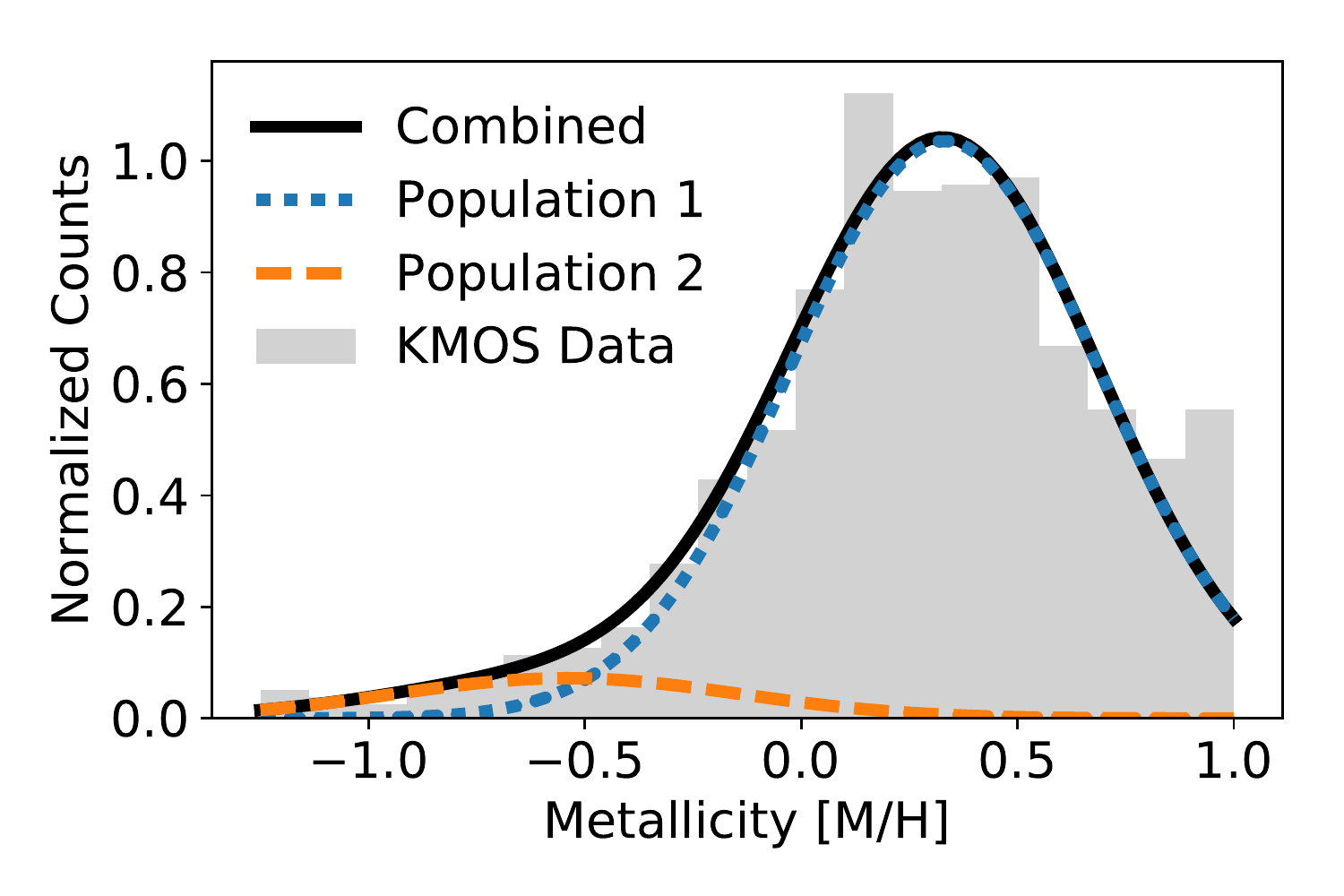}
 \caption{Observed metallicity distribution (filled grey steps) compared with the two populations obtained
with the Gaussian Mixture model, one centered around $[M/H]$=0.3 dex (blue dashed line), and the other peaking at sub-solar metallicity, $[M/H]$ = -0.5 (orange dashed line). About 7\% of the total population are in the sub-solar metallicity component.}
 \label{fig:mh_dist}
 \end{figure}

\begin{figure*}[tbh]
 \center
 \includegraphics[width=6.5in]{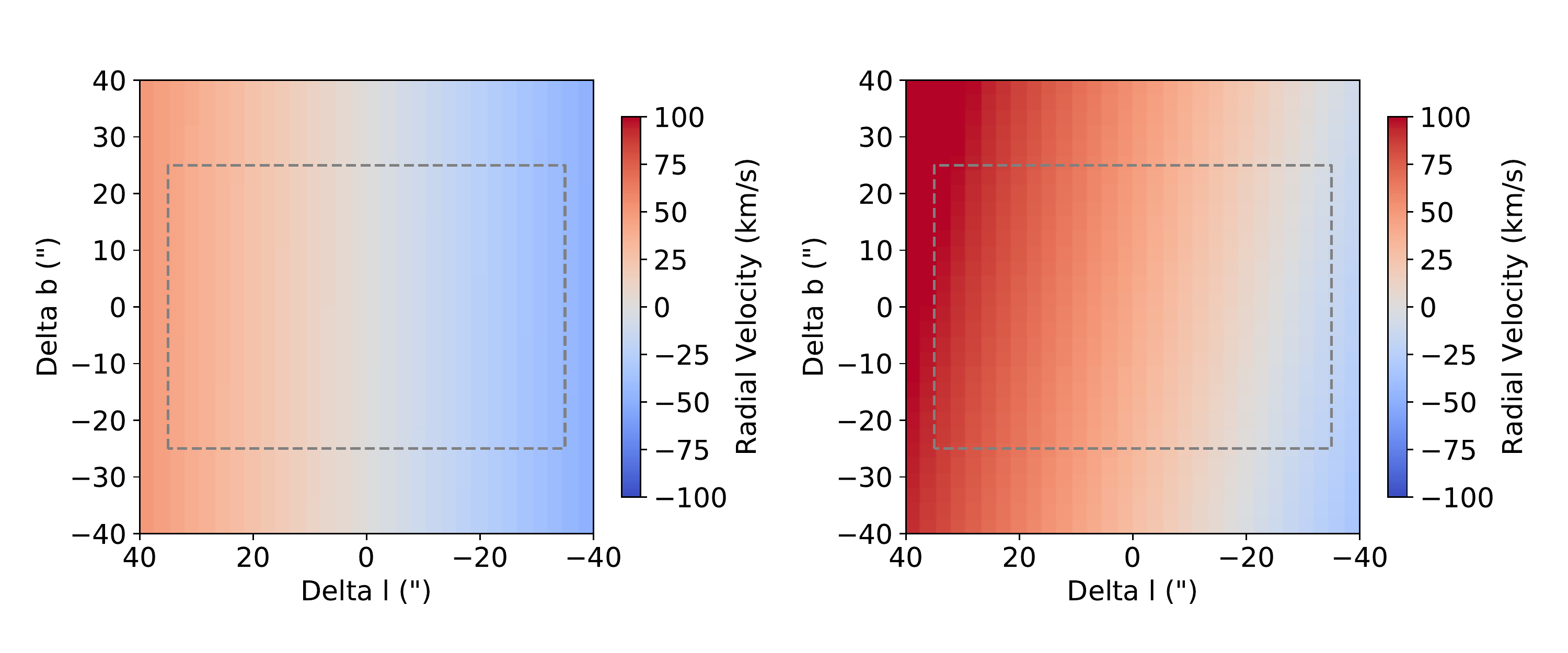}
 \caption{A radial velocity map from the best-fit Gaussian Mixture model of the higher metallicity population (left) and the lower metallicity population (right). The lower metallicity population shows stronger rotation and its rotation axis may be tilted with respect to the higher metallicity population.}
 \label{fig:rotation_map}
 \end{figure*}

We find that there are two distinct populations of stars revealed by the chemo-dynamical model. About 90\% of the stars have a metallicity distribution peaking at $m_1$ = $0.33^{+0.03}_{-0.02}$ dex, while about 7\% of the stars have metallicities peaking at $m_2$ = $-0.54^{+0.29}_{-0.22}$ dex (Fig. \ref{fig:mh_dist}). The parameters for the sub-solar metallicity population have larger uncertainties due to the smaller number of stars in that population. While these two populations have similar velocity dispersions of $\sigma_1 = 91.84^{+2.71}_{-2.71}$ \kms, and $\sigma_2 = 108.96^{+19.37}_{-14.37}$ \kms, their rotational signatures show some distinct differences. The mean velocity of the main component is closer to zero velocity ($v_{z_1} = 0.41^{+4.15}_{-4.55}$ \kms) compared to the positive radial velocity of the sub-solar component ($v_{z_2} = 43.48^{+23.87}_{-22.21}$ \kms). The rotation is also stronger in the sub-solar metallicity component (Fig. \ref{fig:rotation_map}), with a rotation curve that has higher amplitude and is offset from zero velocity at the projected location of the black hole (Fig. \ref{fig:rot_curve}). 
The super-solar metallicity component is rotating slower and its rotation axis is perpendicular to the Galactic plane, while the sub-solar component is rotating faster and has a rotation axis that is slightly tilted. The low number of stars in the sub-solar population do not allow a strong constraint on the orientation of the rotation at this time, with a position angle of 108$^{+45}_{-50}$ degrees. 

To test whether a two-component model is statistically a better fit than a single-component model, we also fit the cluster using a single population model and use Bayesian model selection. Using a 6 parameter single component model, we find the best fit values to be close to the values for Population 1 in the two-component fit. This explains why other studies \citep[e.g.][]{2008A&A...492..419T}, which do not contain metallicity information, are in good agreement with Population 1. However, the Bayes Factor, or the difference in the log evidence between the two models (two-component versus one-component), is about 16. This means that the two-component model is overwhelmingly preferred over the single-component model, even though the two-component model has more model parameters \citep{jeffreys1961}.

\section{Discussion \& Conclusions}
\label{sec:discussion}

\begin{figure}[tbh]
 \center
 \includegraphics[width=3.5in]{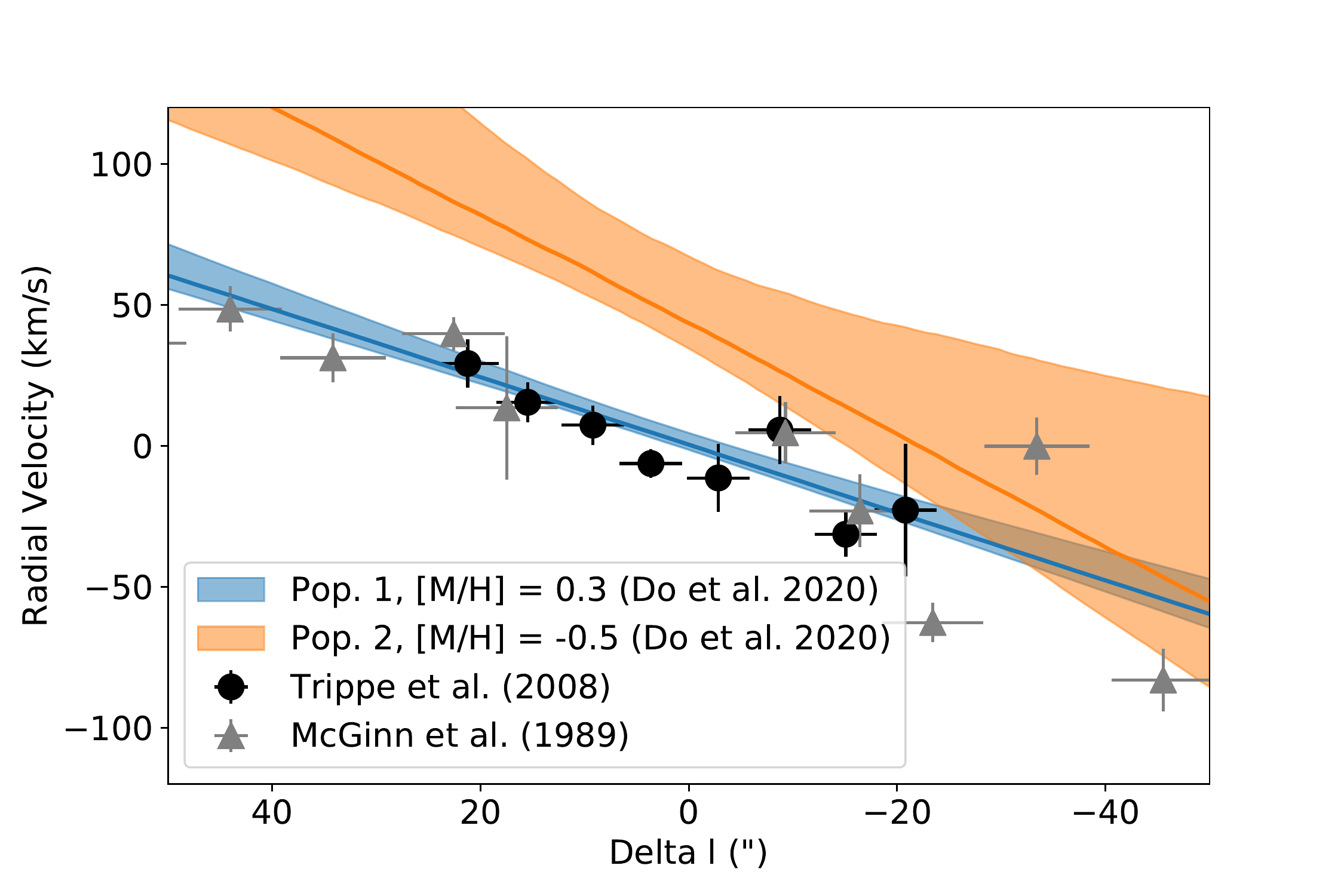}
 \caption{The rotation curves of the two best-fit populations in our model. The higher metallicity population (blue) follows the same rotation curve as has been previously observed (black points), but the lower metallicity population (orange) has a distinctively different rotation curve. It has stronger rotation and may be offset. This suggests that this low metallicity population may be part of a recent infall of a cluster or dwarf galaxy into the Galactic center. }
 \label{fig:rot_curve}
 \end{figure}

By using a chemo-dynamical model, we are sensitive to different components of the MW NSC. The majority of stars in the NSC have super-solar metallicity and rotates in the same direction as the MW disk. This population is consistent with the results of previous studies of the kinematics of the MW NSC. Figure \ref{fig:rot_curve} shows the 1-dimensional rotation curve as a function of Galactic latitude from previous studies compared to the high-metallicity population identified here.

The kinematically distinct low-metallicity population of stars in the MW NSC is identified here for the first time. It is likely that this population was not identified previously because this population consists of only about 10\% of the stars. Previous studies used integrated-light spectroscopy which blended the light of the stars, which would dilute this signal \citep[e.g.][]{1989ApJ...338..824M,2014A&A...570A...2F}. Studies of using individual stars did not include metallicity information, so the signal from this second component was also suppressed \citep[e.g.][]{2008A&A...492..419T,2009A&A...502...91S,2016ApJ...821...44F}. 

We hypothesize that the sub-solar metallicity population may have been accreted into the MW NSC. The differences in metallicity and kinematics may indicate that we are observing the remnants of a disrupted star cluster or dwarf galaxy. Alternatively, in-situ star formation from infalling metal-poor gas may have occurred. Depending on the timescales of relaxation processes \citep[][]{2005PhR...419...65A,2011ApJ...738...99M}, the kinematic signatures may still be distinguishable. The metallicity distribution of about 0.3 dex is rather large for a star cluster; the cores of dwarf galaxies which have more complex star formation history or infalling gas may be more consistent with this spread in metallicity. 

In our companion paper \citet{arcasedda2020}, we use direct N-body simulations to model the infall of a star cluster into an MW-like galactic nucleus \citep[see also][]{2018MNRAS.477.4423A} to place constraints on the possible origin of the observed metal-poor population. Our simulations suggest that the infall of a massive stellar system occurred in between 0.1-3 Gyr ago could give rise to distinguishable kinematic features visible in proper motion and line-of-sight velocity. Our simulations predict that  former star cluster members -- i.e. the metal-poor population -- constitute around 7.3\% of the total stellar population inside 4 pc, in agreement with the observational limits inferred in this work. Comparing models and observations, we conclude that the possible progenitor of the infalling stellar system was either a massive star cluster with mass $10^5-10^7$ $M_\odot$ located ~ 3-5 kpc away from the Galactic center or a dwarf galaxy with mass $\sim10^{10}$ $M_\odot$ initially located at around 100 kpc. 

An alternative explanation may be that the sub-solar metallicity component is not physically located at the Galactic center, and their kinematic signatures are unrelated to the nuclear star cluster. This scenario is not likely because the velocity dispersion of this population is consistent with the main population of stars, indicating the two components likely exist in the same gravitational potential. In addition, their photometry and colors are consistent with the stellar population and extinction at the Galactic center \citep{2010A&A...511A..18S}. 

Future observations will be able to test the hypotheses we present here. The infall of a cluster or dwarf galaxy should leave a stream of stars. This stream may be detected as an anisotropy in the spatial and kinematic distribution of sub-solar metallicity stars compared to super-solar metallicity stars. The abundance ratios of the sub-solar metallicity stars can be used to distinguish between globular cluster-like or dwarf galaxy origins. In addition, our current simple model is limited by the small number of stars. By increasing both the number of stars observed and of the spatial scales of observations, it will be possible to confirm our results and improve the sophistication of dynamical models that can be tested.

\acknowledgements We thank the anonymous referee for helpful comments. We also thank Elena Murchikova, Scott Tremaine, and Denis Erkal for helpful discussions. MAS acknowledges financial support from the Alexander von Humboldt Foundation for the research program ``The evolution of black holes from stellar to galactic scales''. MAS gratefully acknowledges support by the Deutsche Forschungsgemeinschaft (DFG, German Research Foundation) -- Project-ID 138713538 -- SFB 881 (``The Milky Way System''). Part of the simulations presented in this work have been performed with the bwForCluster of the Baden-W\"{u}rttemberg's High-Performance Computing (HPC) facilities thanks to the support provided by the state of Baden-W\"urttemberg through bwHPC and the German Research Foundation (DFG) through grant INST 35/1134-1 FUGG.


\begin{thebibliography}{}
\expandafter\ifx\csname natexlab\endcsname\relax\def\natexlab#1{#1}\fi
\providecommand{\url}[1]{\href{#1}{#1}}
\providecommand{\dodoi}[1]{doi:~\href{http://doi.org/#1}{\nolinkurl{#1}}}
\providecommand{\doeprint}[1]{\href{http://ascl.net/#1}{\nolinkurl{http://ascl.net/#1}}}
\providecommand{\doarXiv}[1]{\href{https://arxiv.org/abs/#1}{\nolinkurl{https://arxiv.org/abs/#1}}}

\bibitem[{{Alexander}(2005)}]{2005PhR...419...65A}
{Alexander}, T. 2005, \physrep, 419, 65, \dodoi{10.1016/j.physrep.2005.08.002}

\bibitem[{{Alfaro-Cuello} {et~al.}(2019){Alfaro-Cuello}, {Kacharov},
  {Neumayer}, {L{\"u}tzgendorf}, {Seth}, {B{\"o}ker}, {Kamann}, {Leaman}, {van
  de Ven}, {Bianchini}, {Watkins}, \& {Lyubenova}}]{2019ApJ...886...57A}
{Alfaro-Cuello}, M., {Kacharov}, N., {Neumayer}, N., {et~al.} 2019, \apj, 886,
  57, \dodoi{10.3847/1538-4357/ab1b2c}

\bibitem[{{Alfaro-Cuello} {et~al.}(2020){Alfaro-Cuello}, {Kacharov},
  {Neumayer}, {Bianchini}, {Mastrobuono-Battisti}, {Luetzgendorf}, {Seth},
  {Boeker}, {Kamann}, {Leaman}, {Watkins}, \& {van de
  Ven}}]{2020arXiv200207814A}
---. 2020, arXiv e-prints, arXiv:2002.07814.
\newblock \doarXiv{2002.07814}

\bibitem[{{Antonini} {et~al.}(2012){Antonini}, {Capuzzo-Dolcetta},
  {Mastrobuono-Battisti}, \& {Merritt}}]{2012ApJ...750..111A}
{Antonini}, F., {Capuzzo-Dolcetta}, R., {Mastrobuono-Battisti}, A., \&
  {Merritt}, D. 2012, \apj, 750, 111, \dodoi{10.1088/0004-637X/750/2/111}

\bibitem[{{Arca-Sedda} {et~al.}(2015){Arca-Sedda}, {Capuzzo-Dolcetta},
  {Antonini}, \& {Seth}}]{2015ApJ...806..220A}
{Arca-Sedda}, M., {Capuzzo-Dolcetta}, R., {Antonini}, F., \& {Seth}, A. 2015,
  \apj, 806, 220, \dodoi{10.1088/0004-637X/806/2/220}

\bibitem[{{Arca-Sedda} \& {Gualandris}(2018)}]{2018MNRAS.477.4423A}
{Arca-Sedda}, M., \& {Gualandris}, A. 2018, \mnras, 477, 4423,
  \dodoi{10.1093/mnras/sty922}

\bibitem[{{Arca-Sedda} {et~al.}(2020){Arca-Sedda}, {Gualandris}, {Do},
  {Feldmeier-Krause}, {Neumayer}, \& {Erkal}}]{arcasedda2020}
{Arca-Sedda}, M., {Gualandris}, A., {Do}, T., {et~al.} 2020, ApJL, accepted

\bibitem[{{Arca-Sedda} {et~al.}(2018){Arca-Sedda}, {Kocsis}, \&
  {Brandt}}]{2018MNRAS.479..900A}
{Arca-Sedda}, M., {Kocsis}, B., \& {Brandt}, T.~D. 2018, \mnras, 479, 900,
  \dodoi{10.1093/mnras/sty1454}

\bibitem[{{Capuzzo-Dolcetta} \& {Miocchi}(2008)}]{2008ApJ...681.1136C}
{Capuzzo-Dolcetta}, R., \& {Miocchi}, P. 2008, \apj, 681, 1136,
  \dodoi{10.1086/588017}

\bibitem[{{Do} {et~al.}(2018){Do}, {Kerzendorf}, {Konopacky}, {Marcinik},
  {Ghez}, {Lu}, \& {Morris}}]{2018ApJ...855L...5D}
{Do}, T., {Kerzendorf}, W., {Konopacky}, Q., {et~al.} 2018, \apjl, 855, L5,
  \dodoi{10.3847/2041-8213/aaaec3}

\bibitem[{{Do} {et~al.}(2015){Do}, {Kerzendorf}, {Winsor}, {St{\o}stad},
  {Morris}, {Lu}, \& {Ghez}}]{2015ApJ...809..143D}
{Do}, T., {Kerzendorf}, W., {Winsor}, N., {et~al.} 2015, \apj, 809, 143,
  \dodoi{10.1088/0004-637X/809/2/143}

\bibitem[{{Do} {et~al.}(2013){Do}, {Lu}, {Ghez}, {Morris}, {Yelda}, {Martinez},
  {Wright}, \& {Matthews}}]{2013ApJ...764..154D}
{Do}, T., {Lu}, J.~R., {Ghez}, A.~M., {et~al.} 2013, \apj, 764, 154,
  \dodoi{10.1088/0004-637X/764/2/154}

\bibitem[{{Do} {et~al.}(2019){Do}, {Hees}, {Ghez}, {Martinez}, {Chu}, {Jia},
  {Sakai}, {Lu}, {Gautam}, {O{\textquoteright}Neil}, {Becklin}, {Morris},
  {Matthews}, {Nishiyama}, {Campbell}, {Chappell}, {Chen}, {Ciurlo},
  {Dehghanfar}, {Gallego-Cano}, {Kerzendorf}, {Lyke}, {Naoz}, {Saida},
  {Sch{\"o}del}, {Takahashi}, {Takamori}, {Witzel}, \&
  {Wizinowich}}]{2019Sci...365..664D}
{Do}, T., {Hees}, A., {Ghez}, A., {et~al.} 2019, Science, 365, 664,
  \dodoi{10.1126/science.aav8137}

\bibitem[{{Feldmeier} {et~al.}(2014){Feldmeier}, {Neumayer}, {Seth},
  {Sch{\"o}del}, {L{\"u}tzgendorf}, {de Zeeuw}, {Kissler-Patig}, {Nishiyama},
  \& {Walcher}}]{2014A&A...570A...2F}
{Feldmeier}, A., {Neumayer}, N., {Seth}, A., {et~al.} 2014, \aap, 570, A2,
  \dodoi{10.1051/0004-6361/201423777}

\bibitem[{{Feldmeier-Krause} {et~al.}(2017){Feldmeier-Krause}, {Kerzendorf},
  {Neumayer}, {Sch{\"o}del}, {Nogueras-Lara}, {Do}, {de Zeeuw}, \&
  {Kuntschner}}]{2017MNRAS.464..194F}
{Feldmeier-Krause}, A., {Kerzendorf}, W., {Neumayer}, N., {et~al.} 2017,
  \mnras, 464, 194, \dodoi{10.1093/mnras/stw2339}

\bibitem[{{Feldmeier-Krause} {et~al.}(2020){Feldmeier-Krause}, {Kerzendorf},
  {Do}, {Nogueras-Lara}, {Neumayer}, {Walcher}, {Seth}, {Sch{\"o}del}, {de
  Zeeuw}, {Hilker}, {L{\"u}tzgendorf}, {Kuntschner}, \&
  {Kissler-Patig}}]{2020arXiv200305998F}
{Feldmeier-Krause}, A., {Kerzendorf}, W., {Do}, T., {et~al.} 2020, arXiv
  e-prints, arXiv:2003.05998.
\newblock \doarXiv{2003.05998}

\bibitem[{{Feroz} {et~al.}(2009){Feroz}, {Hobson}, \&
  {Bridges}}]{2009MNRAS.398.1601F}
{Feroz}, F., {Hobson}, M.~P., \& {Bridges}, M. 2009, \mnras, 398, 1601,
  \dodoi{10.1111/j.1365-2966.2009.14548.x}

\bibitem[{{Fritz} {et~al.}(2016){Fritz}, {Chatzopoulos}, {Gerhard},
  {Gillessen}, {Genzel}, {Pfuhl}, {Tacchella}, {Eisenhauer}, \&
  {Ott}}]{2016ApJ...821...44F}
{Fritz}, T.~K., {Chatzopoulos}, S., {Gerhard}, O., {et~al.} 2016, \apj, 821,
  44, \dodoi{10.3847/0004-637X/821/1/44}

\bibitem[{{Gravity Collaboration} {et~al.}(2019){Gravity Collaboration},
  {Abuter}, {Amorim}, {Baub{\"o}ck}, {Berger}, {Bonnet}, {Brand ner},
  {Cl{\'e}net}, {Coud{\'e} Du Foresto}, {de Zeeuw}, {Dexter}, {Duvert},
  {Eckart}, {Eisenhauer}, {F{\"o}rster Schreiber}, {Garcia}, {Gao}, {Gendron},
  {Genzel}, {Gerhard}, {Gillessen}, {Habibi}, {Haubois}, {Henning}, {Hippler},
  {Horrobin}, {Jim{\'e}nez-Rosales}, {Jocou}, {Kervella}, {Lacour},
  {Lapeyr{\`e}re}, {Le Bouquin}, {L{\'e}na}, {Ott}, {Paumard}, {Perraut},
  {Perrin}, {Pfuhl}, {Rabien}, {Rodriguez Coira}, {Rousset}, {Scheithauer},
  {Sternberg}, {Straub}, {Straubmeier}, {Sturm}, {Tacconi}, {Vincent}, {von
  Fellenberg}, {Waisberg}, {Widmann}, {Wieprecht}, {Wiezorrek}, {Woillez}, \&
  {Yazici}}]{2019A&A...625L..10G}
{Gravity Collaboration}, {Abuter}, R., {Amorim}, A., {et~al.} 2019, \aap, 625,
  L10, \dodoi{10.1051/0004-6361/201935656}

\bibitem[{Jeffreys(1961)}]{jeffreys1961}
Jeffreys, H. 1961, The Theory of Probability, 3rd edn. (Oxford University
  Press)

\bibitem[{{Kerzendorf} \& {Do}(2015)}]{2015zndo.soft28016K}
{Kerzendorf}, W., \& {Do}, T. 2015, {starkit: First real release}, Zenodo
  Source Code Library, doi: 10.5281/zenodo.28016, \dodoi{10.5281/zenodo.28016}

\bibitem[{{Lu} {et~al.}(2013){Lu}, {Do}, {Ghez}, {Morris}, {Yelda}, \&
  {Matthews}}]{2013ApJ...764..155L}
{Lu}, J.~R., {Do}, T., {Ghez}, A.~M., {et~al.} 2013, \apj, 764, 155,
  \dodoi{10.1088/0004-637X/764/2/155}

\bibitem[{{Madigan} {et~al.}(2011){Madigan}, {Hopman}, \&
  {Levin}}]{2011ApJ...738...99M}
{Madigan}, A.-M., {Hopman}, C., \& {Levin}, Y. 2011, \apj, 738, 99,
  \dodoi{10.1088/0004-637X/738/1/99}

\bibitem[{{McGinn} {et~al.}(1989){McGinn}, {Sellgren}, {Becklin}, \&
  {Hall}}]{1989ApJ...338..824M}
{McGinn}, M.~T., {Sellgren}, K., {Becklin}, E.~E., \& {Hall}, D.~N.~B. 1989,
  \apj, 338, 824, \dodoi{10.1086/167239}

\bibitem[{{Neumayer} {et~al.}(2020){Neumayer}, {Seth}, \&
  {Boeker}}]{2020arXiv200103626N}
{Neumayer}, N., {Seth}, A., \& {Boeker}, T. 2020, arXiv e-prints,
  arXiv:2001.03626.
\newblock \doarXiv{2001.03626}

\bibitem[{{Pfuhl} {et~al.}(2011){Pfuhl}, {Fritz}, {Zilka}, {Maness},
  {Eisenhauer}, {Genzel}, {Gillessen}, {Ott}, {Dodds-Eden}, \&
  {Sternberg}}]{2011ApJ...741..108P}
{Pfuhl}, O., {Fritz}, T.~K., {Zilka}, M., {et~al.} 2011, \apj, 741, 108,
  \dodoi{10.1088/0004-637X/741/2/108}

\bibitem[{{Rich} {et~al.}(2017){Rich}, {Ryde}, {Thorsbro}, {Fritz},
  {Schultheis}, {Origlia}, \& {J{\"o}nsson}}]{2017AJ....154..239R}
{Rich}, R.~M., {Ryde}, N., {Thorsbro}, B., {et~al.} 2017, \aj, 154, 239,
  \dodoi{10.3847/1538-3881/aa970a}

\bibitem[{{Ryde} \& {Schultheis}(2015)}]{2015A&A...573A..14R}
{Ryde}, N., \& {Schultheis}, M. 2015, \aap, 573, A14,
  \dodoi{10.1051/0004-6361/201424486}

\bibitem[{{Sch{\"o}del} {et~al.}(2009){Sch{\"o}del}, {Merritt}, \&
  {Eckart}}]{2009A&A...502...91S}
{Sch{\"o}del}, R., {Merritt}, D., \& {Eckart}, A. 2009, \aap, 502, 91,
  \dodoi{10.1051/0004-6361/200810922}

\bibitem[{{Sch{\"o}del} {et~al.}(2010){Sch{\"o}del}, {Najarro}, {Muzic}, \&
  {Eckart}}]{2010A&A...511A..18S}
{Sch{\"o}del}, R., {Najarro}, F., {Muzic}, K., \& {Eckart}, A. 2010, \aap, 511,
  A18+, \dodoi{10.1051/0004-6361/200913183}

\bibitem[{{Trippe} {et~al.}(2008){Trippe}, {Gillessen}, {Gerhard}, {Bartko},
  {Fritz}, {Maness}, {Eisenhauer}, {Martins}, {Ott}, {Dodds-Eden}, \&
  {Genzel}}]{2008A&A...492..419T}
{Trippe}, S., {Gillessen}, S., {Gerhard}, O.~E., {et~al.} 2008, \aap, 492, 419,
  \dodoi{10.1051/0004-6361:200810191}

\end{thebibliography}
\end{document}